\documentclass[a4paper]{spie}  
\usepackage{epsfig}
\usepackage{amssymb}
\usepackage[]{graphicx}

\newcommand{\SigmaHat}{\widehat{\Sigma}}
\newcommand{\taus}{\tau^{\star}}
\newcommand{\taup}{\tau^{\prime}}
\newcommand{\Levy}{L$\acute{\rm e}$vy }

\title{Applications of physical methods in high-frequency futures markets}

\author{M. Bartolozzi\supit{a,b}, C. Mellen\supit{a}, F. Chan\supit{a}, D. Oliver\supit{a}, T. Di
Matteo\supit{c}, T. Aste\supit{c}
 \skiplinehalf
 \supit{a}Research Group, Grinham
Managed Funds, Sydney NSW 2065, Australia \\
\supit{b}Special Research Centre for the Subatomic Structure of
Matter (CSSM), University of Adelaide, Adelaide SA
5005, Australia \\
\supit{c}Department of Applied Mathematics, Research School of
Physical Sciences and Engineering,
 The Australian National University, Canberra ACT 0200, Australia
 }

\authorinfo{Marco Bartolozzi: E-mail: marco.bartolozzi@gmf.com.au}

\begin{document}
 \maketitle

\begin{abstract}
In the present work we demonstrate the application of different
physical methods to high-frequency or tick-by-tick financial time
series data. In particular, we calculate the Hurst exponent and
inverse statistics for the price time series taken from a range of
futures indices.  Additionally, we show that in a limit order book
the {\em relaxation times} of an imbalanced book state with more
demand or supply can be described by stretched exponential laws
analogous to those seen in many physical systems.
\end{abstract}

\keywords{Time series analysis, Hurst exponent, Relaxation times,
Inverse statistics, Limit order book}

\section{Introduction}
\label{sec::introduction}

In recent years the study of financial markets have become more
prevalent among the physics
community\cite{Bouchaud99,Mantegna99,Voit05}. The popularity of
this area of research, also known as econophysics, is partially
due to the large amount of data that has become freely available
to scientists. However, most of the studies deal with sample
periods of the order of days while in fact a large number of
modern financial institutions often deal with high-frequency or
tick-by-tick data.
%
%

The importance of tick-by-tick time series data is certainly
related to the fact that it allows one to address important
questions regarding the dynamics of market microstructure. One
such question relates to the possible presence of higher-order or
non-linear correlations in the price change dynamics, quantities
that can be characterised by calculation of the \emph{Hurst
exponent}~\cite{Feder88}, $H \in [0,1]$,  of the time series. If
non-linear correlations are identified then they can be seen as a
possible source   of price feedback or market \emph{inefficiency}
and as such serve to contradict  the \emph{efficient market
hypothesis} (EMH)~\cite{Dacorogna01}. Such things are very
interesting for   market participants who might, for example, seek
to construct high-frequency trading systems. Related to the
presence of high-order correlations are questions about the nature
of \emph{first passage distribution} of the price time series.
Investigations into the first passage time can be placed within
the framework  of \emph{inverse
statistics}~\cite{Jensen99,Biferale99} and are another tool for
understanding the scaling properties of a times series.
Alternatively, for a high-frequency trader, knowledge of the first
passage distribution is useful in that it can be used to construct
estimates of the \emph{optimal investment horizon}, $\tau$, that
is, the average time required for the price to move by some
specified amount. For Brownian motion, the analytical solution for
$P( \tau )$, the distribution of $\tau$, is
known~\cite{Grimmet94,Ding95,Rangarajan00}. For price evolution
that is non-Brownian (possibly due to high-order correlations) an
empirical model for $P( \tau )$ will need to be derived. Another
set of interesting microstructure questions can be posed around
understanding the dynamics of the limit order book. The limit
order book stores market supply and demand and generally exists in
a state of \emph{disequilibrium}. In fact, there are complex
interplays and feedbacks between the evolution of the order book
state, the order flow in the market, and the evolution of the
market price. Hence arriving at an understanding of the order book
dynamics is important if we are to attempt a complete explanation
of the observed price change dynamics.

In this work we investigate higher-order correlations within
financial time series by calculating the \emph{local} Hurst
exponent, $H_L(t)$, for a range of futures markets. The interest
in the local Hurst exponent is that it allows one to investigate
how scaling within a time series varies in time and with length
scale. For the same markets we investigate also the distribution
of the first passage time and fit an empirical  model to describe
the distribution of the optimal investment horizon $P( \tau )$.
Finally, we study some characteristics of the relaxation of volume
imbalances in limit order books.   Specifically, we look at the
distribution of times that an unbalanced limit order book book
takes to  revert to a zero imbalance or \emph{equilibrium} state.

\section{Local behaviour of Hurst exponents in high-frequency futures contracts}
\label{sec::DFA}

\begin{figure}
\vspace{1cm}
\centerline{\epsfig{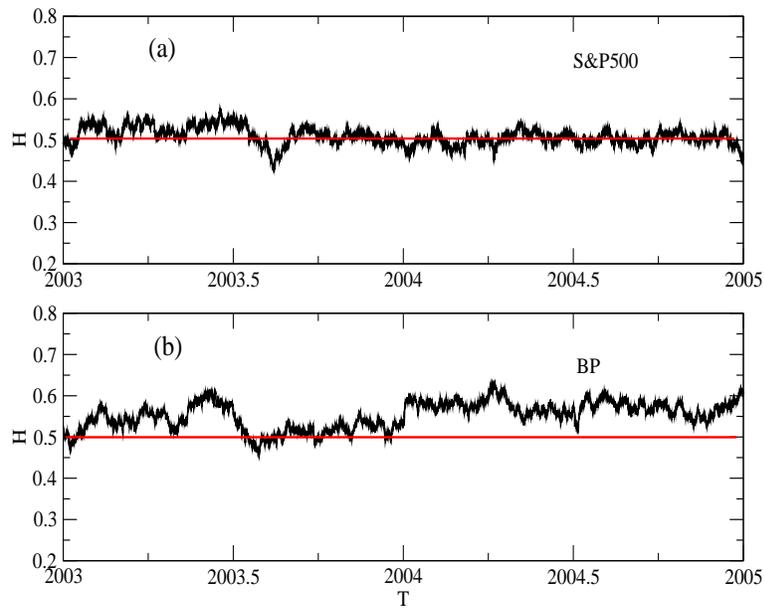}}
\caption{Time series of local Hurst exponents, $H_{L}(t)$, for the
time series S\&P500 (SP)(a) and British Pound exchange rate (BP)
(b) on a scale of approximately 16 days ($L=8192$) and constant
shift $\Delta t=10$ minutes. For the SP, a particularly liquid
market $H$, is always very close to 0.5, irrespective of the
particular period under consideration. The time period goes from
1/1/2003 to 31/12/2004. The BN, instead, displays periods in which
H is significantly different from the random walk benchmark. }
\label{fig::ts-DFA-1}
\end{figure}

In physics, as well as other scientific disciplines, the Hurst
exponent,  $H \in [0,1]$, is often considered as an indicator for
correlations in time series analysis~\cite{Feder88}. In
particular, for $0 \le H < 0.5$ it is said that the behaviour of
the time series is {\em antipersistent}, and conversely, {\em
persistent} for $0.5 < H \le 1$. For completely uncorrelated
movements, as assumed by the ``efficient market hypothesis" of
financial markets, we expect $H=0.5$. As a first step of our
investigation we apply the concept of {\em local Hurst exponent},
$H_{L}(t)$, to different time series of futures contracts sampled
at 1 minute period from 1/1/2003 and ending 31/12/2004. According
to this method, described in Ref.~\cite{Bartolozzi07}, the
exponent $H$ is calculated, via the detrended fluctuation
analysis~\cite{Peng93} (DFA), over a time window of length $L$
much smaller of the length of the entire time series. The
calculation is then repeated by shifting the time window by a
fixed period (in the present analysis 10 minutes) so to obtain an
entire time series of Hurst exponents. An example of the outcome
of this method is given in Fig.~\ref{fig::ts-DFA-1} where the time
series of $H_{L}(t)$ for the S\&P500 (SP) and  the British Pound
exchange rate (BP) futures indices are reported. From the plot, we
can notice how the Hurst exponent is not strictly stationary
during the period under consideration. However for the S\&P500,
Fig.~\ref{fig::ts-DFA-1} (top), the values of $H_{L}(t)$ are
always very close to $0.5$ (apart from a few periods) and,
considering the error over their values, there is no evidence for
long periods of persistency or antipersistency at this temporal
scale. The same does not hold for the BP, Fig.~\ref{fig::ts-DFA-1}
(bottom), where significant deviations from the random walk
behaviour can be observed. Further examples are reported in
Ref.~\cite{Bartolozzi07}.

\begin{figure}
\vspace{1cm}
\centerline{\epsfig{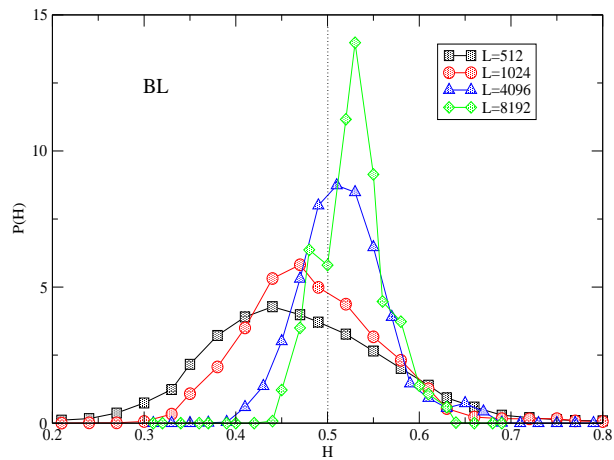}}
\caption{The pdfs for the BOBL (BL) futures show a quite singular
behaviour. The index seems to shift from a slightly persistent to
a slightly antipersistent behaviour as we move toward smaller
scales.} \label{fig::scaling_BL}
\end{figure}

The general statistical behaviour of the local Hurst exponent is
not only sensitive to the particular market but also to the scale
of the observation. In order to highlight these differences we
report in Fig.~\ref{fig::scaling_BL} the pdfs of $H_{L}(t)$ for
various $L$ in the particular case of the BOBL. For this fixed
contract it is evident a change in the average behaviour of the
index: from persistent at large scales to anti-persistent at small
scales.

\begin{figure}
\vspace{2cm} \centerline{\epsfig{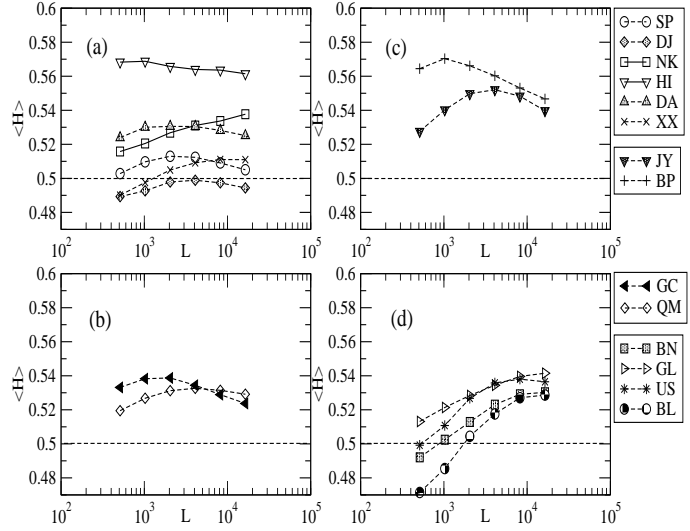}} \caption{Average value of the
local Hurst exponent, $\langle H \rangle_L$, for index futures,
DAX (DA), Euro Stoxx (XX), Standard \& Poor500 (SP), Dow (DJ),
Hang Seng
    (HI) and Nikkei255 (a), commodity futures, Gold COMEX (GC) and Crude Oil E-mini (QM) (b), exchange
rate futures, Japanese Yen (JY) and British Pound
    (BP) (c) and fixed income futures, Eurex Bunds (BN), Long Gilts (GL), Treasury Bonds (US) and BOBL (BL) (d). The horizontal
dotted line is set at 0.5 for visual reference. }
\label{fig::H_vs_L}
\end{figure}

The relationship between the market dynamics and the scale of
observation appears to become more evident when we plot the
average value of the local Hurst exponent, $\langle H \rangle$,
against the scale $L$, Fig.~\ref{fig::H_vs_L}, ranging from 32 to
1 working days approximately. From this graph we infer how time
series belonging to the same sector tend to have a qualitatively
similar scale dependency. The indices futures, for example,
Fig.~\ref{fig::H_vs_L} (a), do not display a strong correlation
between $\langle H \rangle$ and $L$  with the exception of the
Hang Seng (HI) whose persistency increases sharply at smaller
scales. On the other hand a scale dependency is quite evident for
the fixed income products, Fig.~\ref{fig::H_vs_L} (d), where,
interestingly, some time series (BN, US and BL) move from an
antipersistent-like to an persistent-like behaviour as the scale
increases.

As a result of the analysis carried out in this section we can
claim that the actual behaviour of the stock market, described in
terms of Hurst exponent, apart from being influenced by the
particular period of time under consideration and by the maturity
of the
market~\cite{Costa03,DiMatteo03,Cajueiro04,DiMatteo05,Liu07}, is
also related to the scale of observation: this is an extremely
relevant issue for practical applications. In fact, if we consider
long time scales (large $L$), in reality, we are estimating the
{\em average} Hurst exponent over that period.

For more exhaustive discussion on this subject and on ``pitfalls"
of the method with high-frequency data we refer the reader to
Ref.~\cite{Bartolozzi07}.

\section{Optimal exit times in futures indices at tick level}
\label{sec::InvStat}

The framework of {\em inverse statistics} has been recently
introduced in the context of hydrodynamics with the aim of
investigating the scaling properties of the elapsed times between
large fluctuations in the velocity field of turbulent
flows~\cite{Jensen99,Biferale99}.
Given the analogies between the scaling in turbulence and in the
stock market
dynamics~\cite{Dacorogna01,DiMatteo07,Mantegna99,Voit05}, the
application of this technique has been consequently extended to
finance~\cite{Simonsen02,Jensen03,Jensen04,Zhou05,Karpio07}.

The inverse statistics method consists in calculating, for each
element of a time series, the time interval, or {\em exit time},
needed in order to reach for the first time a certain target
level, $R$. For a generic variable $Y$ at time $t$, the exit time,
$\tau_{R} (t)$, can be formally defined as
\begin{equation}
\tau_{R} (t)= {\rm inf} \{ T > 0 \, | \,  Y(t+T)-Y(t) \ge R \}.
\label{eq::def-inverse-stat}
\end{equation}
For clarity of notation, in the rest of the paper we will indicate
the exit time simply as $\tau$.

The {\em first passage distribution}, $P(\tau)$, is then the
probability distribution function (pdf) of the exit times.
In the financial context we can identify the variable $Y(t)$
 with the price of a certain index, $p_r(t)$, or, alternatively, with its
 logarithm. In previous studies~\cite{Simonsen02,Jensen03,Jensen04,Zhou05},
 up to a frequency of 5 minutes, $P(\tau)$
displayed a pronounced peak, or {\em optimal investing horizon},
$\tau^{\star}$, as well as an asymptotical power law decay,
$P(\tau) \sim \tau^{-\alpha}$, with $\alpha \sim 1.5$,
independently on the specific market or sampling period\footnote{
It is worth pointing out that discrepancies from this ``universal"
value have been observed for times shorter then one minute in the
NASDAQ index~\cite{Zhou05}.}. In analogy with the distribution
derived analytically for the Brownian motion~\cite{Simonsen02}
 suggested the following empirical model for the
first passage time distribution:
\begin{equation}
P(\tau)= \frac  {\nu}{\Gamma (\alpha / \nu)} \, \frac{\beta^{2
\alpha}}{(\tau+\tau_{0})^{\alpha+1}} \, {\rm e}^{-\left(
\frac{\beta^2}{\tau+\tau_{0}} \right)^{\nu}},
\label{eq::brownian_inverse_improved}
\end{equation}
where $\nu$, $\beta$ and $\tau_0$ are free parameters.


An interesting and practical issue concerning the first passage
distribution in the stock market is related to the scaling of the
optimal investing horizon, $\tau^{\star}$, with $R$. In a Brownian
market, as expected under the efficient market hypothesis, this
would scale as a power law
\begin{equation}
\tau^{\star} \sim R^{\gamma}, \label{eq::scaling-optimal}
\end{equation}
with exponent $\gamma=2$.
Empirical
studies~\cite{Simonsen02,Jensen03,Jensen04,Zhou05,Karpio07} have
instead revealed, at least for large target returns, a power law
behaviour with $\gamma$ systematically smaller than this benchmark
value and depending on the specific market: in the emerging ones
it results to be significantly smaller than in developed
markets~\cite{DiMatteo05}. However, differences seem to disappear
for data sampled at 5 minutes frequency where the scaling
exponents seem to converge~\cite{Zhou05}.

Given the potential importance in high-frequency trading, in this
section we investigate the statistical properties of the first
passage distribution, $P(\tau)$ for {\em intra-day} price changes,
that is we do not include in the analysis the exit times which
cross the end of day. As target variable for our investigation we
use the straight difference in price, $\tilde{r}_{\tau}(t)=
p_{r}(t+\tau) - p_{r}(t)$ and not the logarithmic returns as in
other studies~\cite{Simonsen02,Jensen03,Jensen04,Zhou05,Karpio07}.
This choice, while providing a more natural unit of measure for
intra-day price changes, enhances the underlying effects related
to the {\em granularity} or discreetness of the fluctuation.

The data used in the analysis are composed of tick-by-tick
snapshots of the order book for the following futures contracts:
DAX (FDX), FTSI-100 (FFI), SPI (2YAP) (stock indices) and Eurex
Bunds (FGBL) (fixed income). These time series span for a period
of 200 working days starting from 2/7/2004 for the FDX and FGBL
and from 13/7/2004 for the 2YAP and the FFI:  for the 2YAP, the
smallest data set, we have approximately $6 \cdot 10^6$ samples
while for the FFI, the largest data set has about $ 2.1 \cdot
10^7$ samples.

\begin{figure}
\vspace{1cm} \centerline{\epsfig{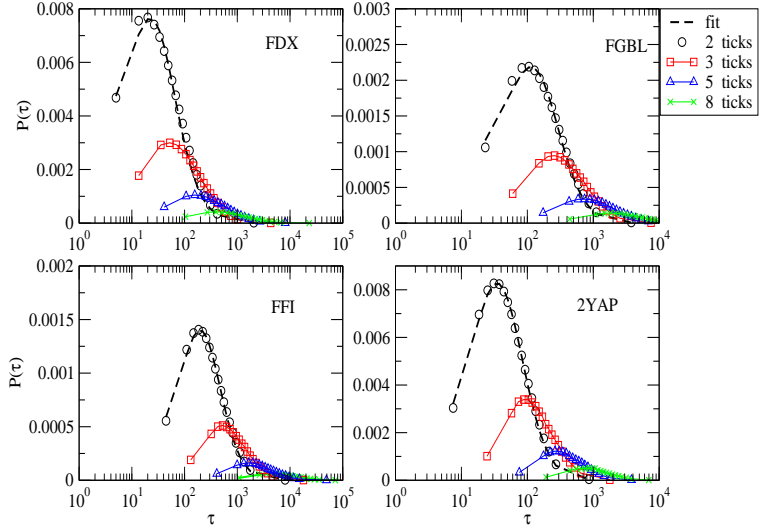}} \caption{First
passage distribution, expressed in tick time, for different
markets and target price displacements (in the legend). The height
of the peak, as well as $\tau^{\star}$, scale with the target
magnitude. For 2 ticks return we also plot the best fit related to
the model of Eq.~(\ref{eq::brownian_inverse_improved}).}
\label{fig::InvStat_global}
\end{figure}

In Fig.~\ref{fig::InvStat_global} we report the first passage
distribution obtained by the analysis of our indices at different
return levels (expressed in ticks).
From the plots it is possible to observe an optimal exit time,
$\tau^{\star}$, that is a peak of the first passage distribution,
getting longer and longer as the target price difference is
increased.
 Fits of the empirical pdfs with the model of
Eq.~(\ref{eq::brownian_inverse_improved}), displayed for a return
of 2 ticks, show  good qualitative agreements. It is important to
stress that what we report is an optimal ``exit tick", rather than
time, during business hours. Since the activity of the market is
not homogeneous during the day, the optimal horizon should be
properly rescaled for practical trading purposes and not used ``as
it is".

We notice also that the optimal exit horizon depends on the
particular market under consideration, as shown in
Fig.~\ref{fig::InvStat_global}. Unfortunately, it is not
straightforward  to identify a relation between $\tau^{\star}$ and
some intrinsic properties of the indices, such as the liquidity
for example. In fact, the Australian 2YAP, despite being a
relatively thin market, has an optimal horizon between the FDX and
the FGBL, which are relatively more liquid.

\begin{figure}
\vspace{1cm} \centerline{\epsfig{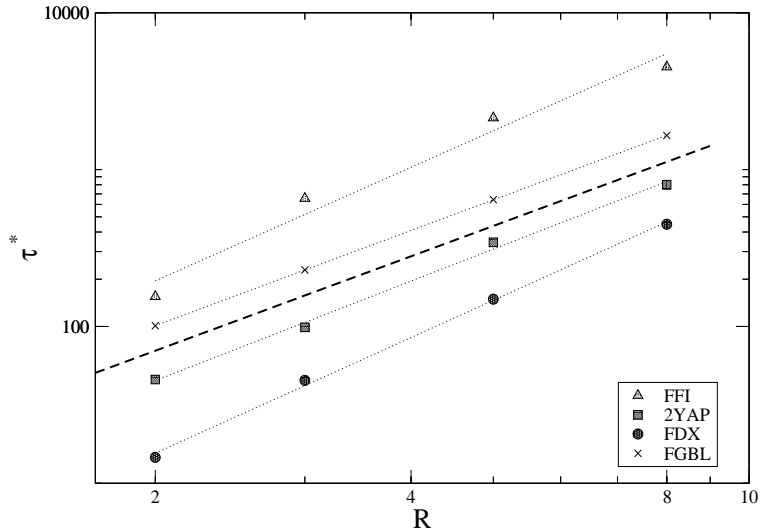}} \caption{Optimal exit
time, $\tau^{\star}$, for the various markets as a function of the
target level. The scales are limited by the intra-day analysis
that we are carrying on. Nevertheless, a power law, as in
Eq.~(\ref{eq::scaling-optimal}), fits quite well the observations
(dotted lines). The dashed line is a power law with exponent 2
plotted for visual comparison.} \label{fig::scaling_maximum}
\end{figure}

\begin{figure}
\vspace{1cm} \centerline{\epsfig{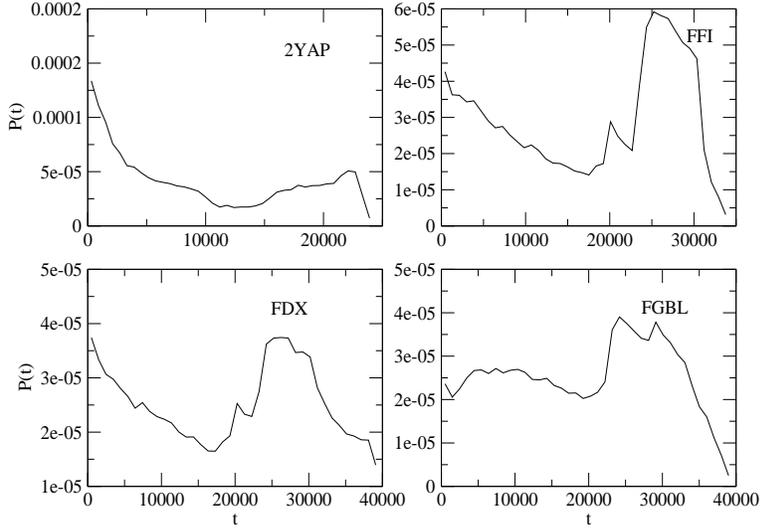}} \caption{Distribution
of entry times for achieving five ticks during a trading day. The
distributions, as expected, follow the market activity during
business hours. The entry time is reported in seconds from the
beginning of the trades.} \label{fig::exit_time}
\end{figure}

From Fig.~\ref{fig::InvStat_global} we notice also that the shape
of the distribution tends to scale with the target price
displacement, as expected also for the Brownian walk theory. In
particular, we focus our attention on the target dependence, such
as the one reported in Eq.~(\ref{eq::scaling-optimal}), of the
exit time $\tau^{\star}$. The results are reported on a double
logarithmic scale in Fig.~\ref{fig::scaling_maximum}.
Despite the limitations of our temporal frame, that is one trading
day, which does not allow to have a particulary broad range of
scales available for analysis, a power law relation cannot be
rejected by the fits, shown in Fig.~\ref{fig::scaling_maximum},
where  the exponents found are $\gamma \gtrsim 2$ for each index.
This result does not match with the ones calculated for regularly
spaced samples~\cite{Simonsen02,Jensen03,Zhou05,Karpio07} where
$\gamma \lesssim 2$. However, there are very important differences
in the investigation lines other that the data sets used. In
particular, we consider tick-by-tick data as opposed to evenly
sampled  and we limit ourselves to a single business day. It is
worth pointing out that the values of the exponents found in the
present analysis, which would indicate a subdiffusive behaviour
for the price moments, have to be considered in the light of the
non-homogeneous activity of the market during the day. We refer,
in particular, to the series of zero returns that are more likely
to be realized during the lunch hours and that can definitely
influence the results. To stress this fact, we report in
Fig.~\ref{fig::exit_time}, the distribution of the daily entry
times necessary to achieve a return of 5 ticks. As expected, these
results closely relate to the daily trading activity with a valley
in the middle of the day where the frequency of the data arrival
gets lower. Moreover, the distributions of entry times result to
be independent on the target tick.

\section{Relaxation times in the limit order book}
\label{sec::relaxation_times}

The study of limit order books\footnote{The order book, or limit
order book, represents the ensemble of all the limit sell and buy
orders which are present in the market at certain time. Each order
is characterized by its limit price, that is the price at which it
has to be executed, and its volume. What it is usually referred as
``price", or ``mid point price", of a stock, $p_r$, is the mean
value between the price at the best ask and the best bid in the
book, that is a simple transform of the demand and the supply.}
has recently attracted the attention of different econophysics
groups around the
world~\cite{Farmer03,Lillo04,Lillo05,Weber05,Weber06,Bouchaud06}.
Understanding its complex mechanisms implies an understanding of
the dynamics of the stock market at a microscopic level, that is,
using a metaphor borrowed from physics, an understanding from
``first principles". Far from being so ambitious we investigate
the statistical {\em relaxation} properties of the aggregated
demand and supply for two futures indices: the German DAX (FDX)
and the Australian SPI (2YAP). The FDX data are available form
20/07/2004 to 08/12/2006 for a total of $\sim 5 \cdot 10^{7}$
samples. Despite the 2YAP data set is one week longer (starting on
the 13/07/2004) the number of samples in this case is ``just"
$\sim 6\cdot 10^{6}$ reflecting a lower frequency of updating. In
particular, we refer to relaxation in terms of {\em volume
imbalance}, $\Sigma$, defined as
\begin{equation}
\Sigma(t)=\sum_{i=1}^{N_{b}} V_{i}^{b}(t)-
\sum_{i=1}^{N_a}V_{i}^{a}(t), \label{eq::imbalance}
\end{equation}
where $V_{i}^{a,b}$ is the total volume at the $i$th level in the
ask ($a$) or bid ($b$) side, $N_{a,b}$ the total number of levels
on the respective side of the book\footnote{Given the limitation
of our database, in the present study $N_{a}=N_b=10$ for the FDX
and 5 for the 2YAP. However, given that most of the relevant
information is contained in the first levels~\cite{Cao03}, we
believe that we have a significantly faithful representation of
the overall demand and supply.} and $t$ the tick time\footnote{The
definition of volume imbalance given in Eq.~(\ref{eq::imbalance})
takes into account just the total number of orders in the book
without considering their ``spatial" distribution relative to the
mid-point price, for example. The distribution of the volumes can,
indeed, be of great importance and plays a major role in the
appearance of large price
fluctuations~\cite{Farmer03,Lillo05,Weber05,Weber06,Alvarez91,Bouchaud02}.
However, in the present work we want to keep the discussion as
simple as possible and, therefore, we focus on the homogeneous
case.}.
However, the above definition presents a drawback, that is
 the average order size can differ from market to market and,
therefore, the value of $\Sigma$ should be rescaled in order to
standardize the results. In order to avoid this inconvenience we
use the {\em normalized volume imbalance}, $\SigmaHat$, defined
between [-1,1] according to
\begin{equation}
\SigmaHat(t)=\frac{\sum_{i=1}^{N_{b}} V_{i}^{b}(t)-
\sum_{i=1}^{N_a}V_{i}^{a}(t)}{\sum_{i=1}^{N_{b}} V_{i}^{b}(t)+
\sum_{i=1}^{N_a}V_{i}^{a}(t)}. \label{eq::imbalance_norm}
\end{equation}

\begin{figure}
\vspace{1cm} \centerline{\epsfig{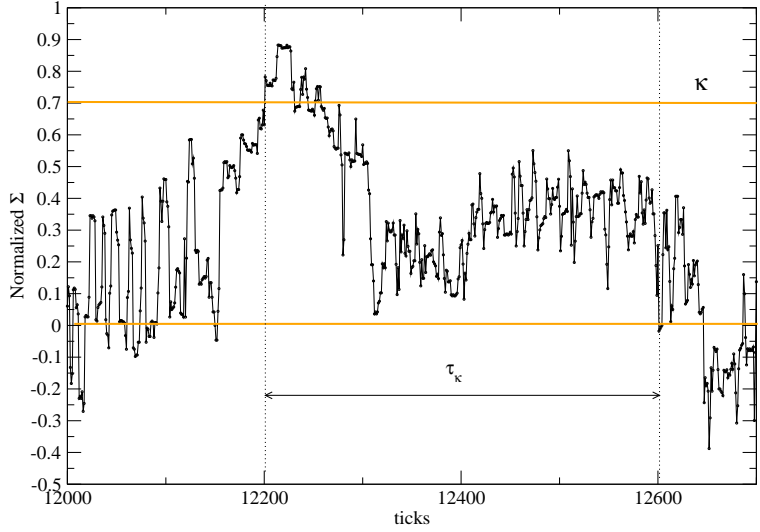}}
\caption{Propaedeutic plot for the definition of $\tau_{\kappa}$
given a certain threshold ($\kappa =0.7$ in orange). The time
series of $\SigmaHat$ is expressed in tick time. The trade time is
calculated by evaluating the number of trades between entry and
exit.} \label{fig::propaedeutic_2YAP}
\end{figure}

The {\em relaxation time} for a certain {\em threshold} volume
imbalance $\kappa$ is then defined as
\begin{equation}
\tau(t_{\kappa})= {\rm inf} \{ T > 0 \, | \,
\SigmaHat(t_{\kappa}+T) \cdot \SigmaHat(t_{\kappa}) \le 0 \},
\label{eq::def-tau}
\end{equation}
given the entry time $t_{\kappa}$,
Fig.~\ref{fig::propaedeutic_2YAP}, belonging to the set $\{ t_k
\}$ defined as
\begin{eqnarray}
\{ t_{\kappa} \} = & \{ t > 0 \, | \,  (| \SigmaHat(t)| - \kappa)
\cdot  (|\SigmaHat(t-1)| - \kappa)  <0 \wedge \cdots \nonumber \\
& (| \SigmaHat(t)| - \kappa) -  (| \SigmaHat(t-1)| - \kappa)
>0 \}. \label{eq::def-entry-time}
\end{eqnarray}

In order to simplify the notation, from now on, we use the
convention $\left\{ \tau_{\kappa} \right\}\equiv \left\{
\tau(t_{\kappa})\right\}$.

Relaxation times for the order book are very important from both a
theoretical point of view and for the purposes of technical
trading. In the first case a proper investigation would be able
shed some light on the dynamical behaviour which characterize the
traders reaction to a particular condition of the market, that is
when there is imbalance between demand and supply. This
``out-of-equilibrium" condition can be, in principle, very
different from the situation when the book is balanced: the
question is, how do the market participants act to drive the
market to a new (meta-stable) ``equilibrium"? In the context of
technical trading, relaxation times could be used to determine
temporal thresholds for the closure of a trade that was
conditioned to a certain book configuration.

\begin{figure}
\vspace{1cm} \centerline{\epsfig{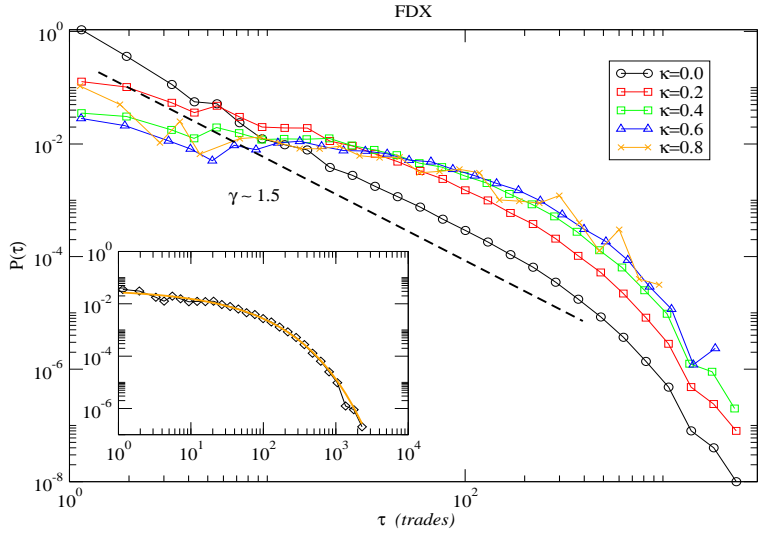}} \caption{
Probability distribution function, for FDX, of relaxation times,
$\tau_{\kappa}$, expressed in trades, for different $\kappa$
values. In the inset we show the fit, for $\SigmaHat=0.4$, against
the stretched exponential of Eq.~(\ref{eq::stretched_exp}). This
distribution provides a very good description of the empirical
pdf. The dashed line, instead, is a power law displayed for visual
comparison. } \label{fig::relaxOB_trades_FDX_all}
\end{figure}

\begin{figure}
\vspace{1cm} \centerline{\epsfig{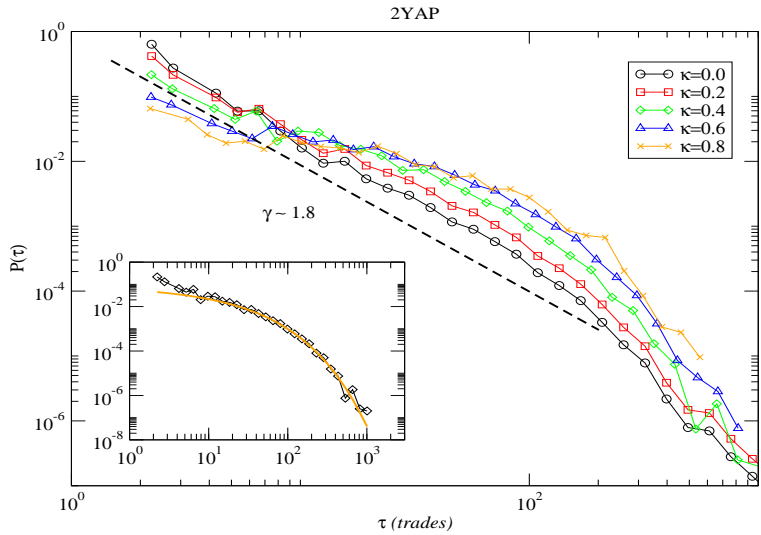}}
\caption{ Same plot as Fig.~\ref{fig::relaxOB_trades_FDX_all} but
for the 2YAP. Also in this case the stretched exponential fit,
shown in the inset, seems to provide a good model for the
distribution of relaxation times.}
\label{fig::relaxOB_trades_2YAP_all}
\end{figure}

In Figs.~\ref{fig::relaxOB_trades_FDX_all} and
\ref{fig::relaxOB_trades_2YAP_all}, respectively for the FDX and
2YAP, we report the empirical probability distribution functions
(pdfs), $P$, of the relaxation
 times, $\tau_{\kappa}$ for various thresholds $\kappa$ of the normalized volume imbalance.
From these plots we can infer a clear correlation between the
shape of the distribution and the threshold imbalance: as we move
from large to small $\kappa$s the pdfs change in an almost
continuous fashion.
Moreover, we find that over almost all the range of $\kappa$ under
examination\footnote{For $\kappa \gtrsim 0.7$ we do not have many
samples available and therefore the statistics results to be more
ambiguous.} the distributions are very well described by a
stretched exponential,
\begin{equation}
P(\tau) \propto {\rm e}^{-\left( \frac{\tau}{\tilde{\tau}}
\right)^{\alpha}}, \label{eq::stretched_exp}
\end{equation}
where $\tilde{\tau}$ is the characteristic time scale and the
exponent $\alpha$ belongs to the interval (0,1] . The insets of
Figs.~\ref{fig::relaxOB_trades_FDX_all}
and~\ref{fig::relaxOB_trades_2YAP_all} report the fits for
$\kappa=0.4$. These findings share interesting similarities with
relaxations in solid state
physics~\cite{Alvarez91,Weron96,Magdziarz06} where it has been
shown that different models can explain the rise of stretched
exponential distributions. In particular, the convolution of
Poisson processes with different characteristic times seems to be
one of the most plausible
scenarios~\cite{Klafter86,Alvarez91}\footnote{Convolution of
distributions, such as the one in Eq.(\ref{eq::convolution_exp}),
is common also in other areas of physics as, for example,
hydrodynamic turbulence where it has been used to interpret the
asymptotic power law decay of the distribution of the fluctuations
in radial velocity fields of turbulent
flows~\cite{Castaing90,Beck03}. }.
In the continuous limit, we can formalize this concept through the
the Kohlrausch-Williams-Watts equation (KWW),
\begin{equation}
{\rm e}^{-\left( \frac{\tau}{\tilde{\tau}} \right)^{\alpha}}=
\int_{0}^{\infty} {\rm e}^{-\frac{\tau}{\taup}} \, \rho(\taup)
 \, d\taup \label{eq::convolution_exp}
\end{equation}
where $\rho(\taup)$ is the distribution of independent time
scales, or {\em Debye distribution}, and can be calculated
numerically for a fixed value of
$\alpha$~\cite{Alvarez91}\footnote{The integration needed to
obtain Debye densities can be numerically troublesome. Some
algorithms, such as the maximum a posteriori (MAP) and
regularization methods, are discussed in~\cite{Honerkamp98}.}. The
parameters of the  stretched exponential, therefore, are related
to the specific form of the Debye distribution.
Despite its intuitive interpretation, this framework presents a
drawback. In fact, assuming that Eq.(\ref{eq::convolution_exp})
holds, we have to postulate the existence of a class of Debye
distributions that can justify the ubiquity of stretched
exponential relaxations.
This issue can be partially solved by realizing that a \Levy
stable distribution $L_{\alpha,\beta}$ with parameters $\alpha \in
(0,1)$ and $\beta=-1$ is a solution of
Eq.~(\ref{eq::convolution_exp})~\cite{Paul99}. The parameters
characterize, respectively, the ``fatness" of the distribution's
tails and the degree of asymmetry and, therefore, $\beta=-1$
corresponds to a positively defined support for
$L_{\alpha,\beta}$. The strength of this framework
 relies on the fact that  \Levy distributions are attractors in the
distributions  space and, therefore, the appearance of a stretched
exponential would be justified in the standard framework of
stochastic processes~\cite{Paul99}. In particular, \Levy
distributions are fixed points for the addition of i.i.d. random
variables, $x_i$, whose individual distributions are
asymptotically a power law, $P_{i}(x_{i}) \sim |x_{i}|^{-1-\nu}$
with $0<\nu<2$, that is, with infinite variance.

Coming back to the relaxation times observed for the volume
imbalance, Figs.~\ref{fig::relaxOB_trades_FDX_all} and
\ref{fig::relaxOB_trades_2YAP_all}, and given the previously
mentioned conditions for the appearance of \Levy distributions, we
can argue that such an attractor for $\rho(\taup)$ in
Eq.(\ref{eq::convolution_exp}) can result as a consequence of the
``fat" tails intrinsic in the human
activity~\cite{Barabasi05,Bartolozzi05,Goh06} and, therefore, help
justify the empirical results\footnote{The relevance of \Levy
distributions in finance has been pointed out by different
previous studies~\cite{Bouchaud99,Mantegna99,Paul99,Voit05}. In
particular, \Levy ``truncated" distributions have been claimed to
fit considerably well the empirical distribution of price
fluctuations over different temporal scales.}. However, we still
have to understand how the changes in $\rho$ are related to the
changes in $\kappa$.
One possible scenario involves the set of ``herding
effects"~\cite{Cont00,Bartolozzi04}. In this case, as the
imbalance in the book increases (hence $|\kappa| > 0$), well
informed individuals and institutions (here, those with knowledge
of the instantaneous order book state) invoke order submission and
cancellation strategies that are more adapted to the large
observed imbalance - for example, strategies that require longer
waiting times between limit order submission and cancellation, or
the submission of limit orders further from the current mid point
price.
Thus, as the informed market participants tend towards a broadly
similar set of strategies, the heterogeneity within the market is
reduced, leading to changes in the Debye distribution of order
arrival and cancellation waiting times and, eventually, to changes
in the distribution of imbalance relaxation times.
Of course this scenario is speculative and it would be important
to back up these qualitative arguments with rigorous numerical
simulations. Moreover, numerical calculations of Debye
distributions could help to shed some light on this issue.

 For $\kappa \sim 0$ the situation is slightly
different. In this regime the distributions become so stretched
that a power law, $P(\tau) \propto \tau^{-\gamma}$, provides a
better description of them in terms of least squared error. The
change in the shape of the distributions, from stretched
exponential towards a power law can still be interpreted in the
framework of Eq.~(\ref{eq::convolution_exp}) given that the
convolution of exponentials can give rise to asymptotic power
laws~\cite{Sornette04}. More intuitively, we can interpret this
``near-equilibrium"  state in terms of diffusion processes. In
fact, without any extra information given by the difference in
demand and supply new limit orders are distributed on each side of
the book according to a large number of heterogeneous strategies
based on other indicators. This phase implies a diffusion process
for the variable $\Sigma$ (or $\SigmaHat$) and the appearance of
power law relaxation times identified with a problem of {\em first
return to the origin}~\cite{Sornette04}. In this framework, for a
fractional Brownian motion it has been analytically shown that the
asymptotic pdf of waiting times between two consecutive crossings
of the origin would be a power law with exponent, ${H-2}$ where
$H$ is the Hurst parameter of the walker~\cite{Ding95}. Note that
for the simple Brownian motion $H=0.5$ and, therefore, the
asymptotic power law exponent would be
$\gamma=1.5$~\cite{Grimmet94,Ding95,Rangarajan00}.
For the time series under examination
 the exponent of the power law estimated for small imbalances,
 $\kappa \sim 0$, is $\gamma \sim 1.5$ for the FDX and $\gamma \sim  1.8 $ for the 2YAP.
 A dashed line for visual guidance is reported in Figs.~\ref{fig::relaxOB_trades_FDX_all}
 and~\ref{fig::relaxOB_trades_2YAP_all}.
 The difference in the two exponents can be justified, at least,
  by two factors: difference in liquidity (the FDX is more
 liquid compared to the 2YAP and, as a consequence, its behaviour is closer to that of a
 random walk) and lack of information
   for the 2YAP (we have just 5 levels on each size of the book and
  more information may be ``hidden" at deeper levels).

\begin{figure}
\vspace{1cm} \centerline{\epsfig{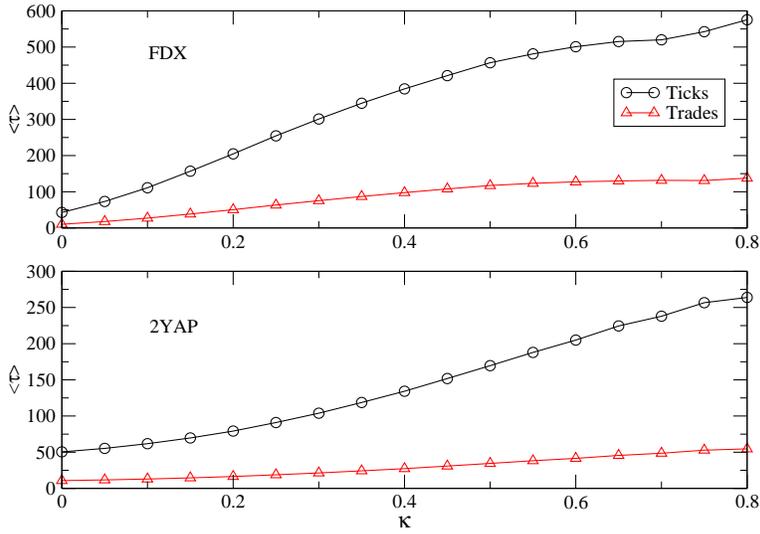}} \caption{Mean relaxation
times expressed in trades and ticks for the FDX (top) and 2YAP
(bottom). In the mid range of imbalances, the increase of the
average price can be well described by a linear function of
$\kappa$.} \label{fig::meanTimeRev0}
\end{figure}

As a final note, we report in Fig.~\ref{fig::meanTimeRev0} the
estimate of mean relaxation times as a function of $\kappa$ in
both trade time and tick time. For $ 0.1 \lesssim \kappa \lesssim
0.6$, the increase of the mean relaxation time is a linear fiction
of $\kappa$ and, given that for a stretched exponential
distribution the average relaxation time is given by
\begin{equation}
\langle \tau \rangle = \frac{\tilde{\tau}}{\alpha} \Gamma \left(
\frac{1}{\alpha}  \right),
\end{equation}
then we can infer that, in this regime, $\tilde{\tau}/\alpha
\propto \kappa$.

\section{Conclusion}

This work has focussed on demonstrating the application of a range
of physical methods to high-frequency and tick-by-tick financial
time series. The results presented relate to the local
time-varying scaling properties of the price time series, the
distribution of the so-called optimal investment horizon and also
to the relaxation properties of disequilibrium limit order books.

In section~\ref{sec::DFA} we demonstrated the use of the concept
of local Hurst exponent in order to investigate the short scale
dynamical properties of the correlations in different futures
contracts (indices, commodities, exchange rate and fixed income)
from the beginning of 2003 to the end of 2004.

Analysis of the behaviour of $H_{L}(t)$ at different scales, and
in particular its distribution, points out a scale dependent and
non-stationary evolution of this scaling exponent, independent of
the specific kind of contract. The Eurex Bunds, BOBL and U.S Treasury Bonds, for example, display
an average persistent behaviour over time scales of approximately
three weeks but they then become antipersistent, on average, for
time scales of the order of one day. Moreover, we observe changes
in the shape of the pdfs of $H_{L}(t)$ with time. This fact points
to the existence of different market phases in the two years
period from 1/1/2003 to 31/12/2004 and, therefore, evidence for
non-stationarity.
These empirical facts are in contrast with the EMH hypothesis,
according to which $H_{L}(t)$ should be constant and equal to 0.5
for each time scale. It is worth to stress that the dynamical behaviour of the Hurst
exponent is not related only to the liquidity of the market but
also to the variety of time horizons involved in the trade of a
particular asset. As a consequence, markets which involve many
exogenous agents, such as the S\&P500, tend to be more
``efficient".
A point we would like to emphasise is that the concept of Hurst
exponent for non-stationary time series has a practical validity
only in the period and the scale of observation. By estimating $H$
with a large sample, due to the coarse-grain procedure of the DFA
algorithm, we lose the local information and we obtain an
``average" value over that period. This can or cannot be a problem
for technical trading: it depends on the horizon we are interested
in.

In section~\ref{sec::InvStat} we have shown the existence of an
optimal exit time, or more correctly ``exit tick", necessary to
achieve a pre-fixed return. Moreover, even if on a limited range,
the scaling of this quantity with the target tick can be well
approximated with a power law with exponent $\gamma  \gtrsim 2$
depending on the market under consideration. Previous studies have
revealed a similar qualitative behaviour but with significant
discrepancies in the scaling index
value~\cite{Simonsen02,Jensen03,Zhou05,Karpio07}. The
discrepancies can be taken into account by the tick nature of our
data and the restriction of our analysis to business hours.
To the best of our knowledge it is the first time that the optimal
exit horizon has been calculated at tick-by-tick frequency {\em
inside} a trading day. From the practitioner point of view the
existence of such a stylized fact can potentially be exploited but
its value has to be taken with care. In fact, the tick frequency
depends on the intra-day pattern activity of the market and,
therefore, the estimated value of $\taus$. Note the usage of real
time, that is sampling at a regular frequency, will lead to
similar problems.

In section~\ref{sec::relaxation_times} we reported on
investigations into the dynamical properties of the order book
conditioned to an imbalance in the aggregated demand/supply. In
particular, we focus on the time required for the order book to
reach a new (meta-stable) ``equilibrium". To the best of our
knowledge this is the first investigation of this kind.
Empirical results show how the relaxation times behave in a way that is analogous to
that found in certain physical systems. In particular, the
distribution $P(\tau_{\kappa})$ seems to have a relatively broad
spectrum that can be very well described by a stretched
exponential distribution with characteristic parameters that
depend on the threshold imbalance. This dependence can be related
to a decrease in the heterogeneity of trader strategies as we move
towards larger imbalances in the book, for example, due to herding
mechanisms. For small initial imbalances, instead, diffusion
processes seem to dominate the dynamics of the book.

In conclusion, we have shown that methods of physical analysis
produce useful and potentially valuable insights into the dynamics
of market microstructure. A feature of the results presented in
this work is that they would not have been obtained without access
to a high-frequency or tick-by-tick record of both market trade
data and the limit order book state. We note that econophysicists
and market practitioners are increasingly turning to the study of
market microstructure, with the aim of gaining a better
understanding of the \emph{complex system} that is the financial
marketplace. Still, questions on items such the source of market
inefficiencies and the nature of the complex interplay between the
limit order book, price evolution and order flow remain largely
unaddressed. It is for these reasons that we expect to see the
types of methods and time series discussed in this work to begin
to feature ever more prominently in future econophysics
literature.

\section*{Acknowledgement}
The authors would like to acknowledge Matt Fender for his precious
IT support and Richard Grinham for fruitful discussions. T.D.M.
and T. A. acknowledge partial support from COST P10 ``Physics of
Risk" project and ARC Discovery Projects: DP03440044 (2003) and
DP0558183 (2005).

\bibliography{SPIE2007III}

\begin{thebibliography}{10}

\bibitem{Bouchaud99}
J.-P. Bouchaud and M.~Potters, {\em Theory of financial risk}, Cambridge
  University Press, Cambridge, 1999.

\bibitem{Mantegna99}
R.~N. Mantegna and H.~E. Stanley, {\em An introduction to econophysics:
  correlation and complexity in finance}, Cambridge University Press,
  Cambridge, 1999.

\bibitem{Voit05}
J.~Voit, {\em The Statistical Mechanics of Financial Markets}, Spriger-Verlag,
  Berlin, 2005.

\bibitem{Feder88}
J.~Feder, {\em Fractals}, Plenum Press, New York, 1998.

\bibitem{Dacorogna01}
M.~M. Dacorogna, R.~Gen\c{c}ay, U.~M\"uller, R.~B. Olsen, and O.~V. Pictet,
  {\em An introduction to high-frequency finance}, Academic Press, San Diego,
  2001.

\bibitem{Jensen99}
M.~H. Jensen, ``Multiscaling and structure functions in turbulence: An
  alternative approach,'' {\em Phys. Rev. Lett.}~{\bf 83}, pp.~76--79, Jul
  1999.

\bibitem{Biferale99}
L.~Biferale, M.~Cencini, D.~Vergni, and A.~Vulpiani, ``Exit time of turbulent
  signals: A way to detect the intermediate dissipative range,'' {\em Phys.
  Rev. E}~{\bf 60}, pp.~R6295--R6298, Dec 1999.

\bibitem{Grimmet94}
G.~Grimmett and D.~Stirzaker, {\em Probability and Random Processes}, Oxford
  University Press, New York, 1994.

\bibitem{Ding95}
M.~Ding and W.~Yang, ``Distribution of first return time in fractional
  {B}rownian motion and its application to the study of on-off intermittency,''
  {\em Physical Review E}~{\bf 52}, p.~207, 1995.

\bibitem{Rangarajan00}
G.~Rangarajan and M.~Ding, ``First passage time distribution for anomalous
  diffusion,'' {\em Physics Letters A}~{\bf 273}, pp.~322--330, 2000.

\bibitem{Bartolozzi07}
M.~Bartolozzi, C.~Mellen, T.~Di~Matteo, and T.~Aste, ``Multi-scale correlations
  in different futures markets,'' {\em The European Physical Journal B}~{\bf
  58(2)}, pp.~207--220, 2007.

\bibitem{Peng93}
C.-K. Peng, S.~V. Buldyrev, S.~Havlin, M.~Simons, H.~E. Stanley, and A.~L.
  Goldberger, ``Mosaic organization of {DNA} nucleotides,'' {\em Physical
  Review E}~{\bf 49}, pp.~1685--1689, Feb 1994.

\bibitem{Costa03}
R.~L. Costa and G.~L. Vasconcelos, ``Long-range correlations and
  nonstationarity in the {B}razilian stock market,'' {\em Physica A}~{\bf 329},
  p.~231, 2003.

\bibitem{DiMatteo03}
T.~Di~Matteo, T.~Aste, and M.~M. Dacorogna, ``Scaling behaviors in differently
  developed markets,'' {\em Physica A}~{\bf 324}, p.~183, 2003.

\bibitem{Cajueiro04}
D.~O. Cajueiro and M.~B. Tabak, ``The {H}urst exponent over time: testing the
  assertion that emerging markets are becoming more efficient,'' {\em Physica
  A}~{\bf 336}, p.~231, 2004.

\bibitem{DiMatteo05}
T.~Di~Matteo, T.~Aste, and M.~M. Dacorogna, ``Long term memories of developed
  and emerging markets: using the scaling analysis to characterize their stage
  of development,'' {\em Journal of Banking \& Finance}~{\bf 29}, p.~827, 2005.

\bibitem{Liu07}
R.~Liu, T.~Di~Matteo, and T.~Aste, ``True and apparent scaling: The proximities
  of the {M}arkov-switching multifractal model to long-range dependence,'' {\em
  Physica A}~{\bf 383}, p.~35, 2007.

\bibitem{DiMatteo07}
T.~Di~Matteo, ``Multi-scaling in finance,'' {\em Quantitative Finance}~{\bf
  7(1)}, p.~21, 2007.

\bibitem{Simonsen02}
I.~Simonsen, M.~H. Jensen, and A.~Johansen, ``Optimal investment horizon,''
  {\em The European Physical Journal B}~{\bf 27}, pp.~583--586, 2002.

\bibitem{Jensen03}
M.~H. Jensen, A.~Johansen, and I.~Simonsen, ``Inverse statistics in economics:
  the gain-loss asymmetry,'' {\em Physica A}~{\bf 324}, p.~338, 2003.

\bibitem{Jensen04}
M.~H. Jensen, A.~Johansen, F.~Petroni, and I.~Simonsen, ``Inverse statistics in
  the foreign exchange market,'' {\em Physica A}~{\bf 340}, p.~678, 2004.

\bibitem{Zhou05}
W.-X. Zhou and W.-K. Yuan, ``Inverse statistics in stock markets: universality
  and idiosyncracy,'' {\em Physica A}~{\bf 353}, p.~423, 2005.

\bibitem{Karpio07}
K.~Karpio, M.~A. {Zaluska–-Kotur}, and A.~Orlowski, ``Gain-loss asymmetry for
  emerging stock markets,'' {\em Physica A}~{\bf 375}, p.~599, 2007.

\bibitem{Farmer03}
J.~D. Farmer, L.~Gillemot, F.~Lillo, S.~Mike, and A.~Sen, ``What really causes
  large price changes?,'' {\em Quantitative Finance}~{\bf 4(4)}, pp.~383--397,
  2004.

\bibitem{Lillo04}
F.~Lillo and M.~J.~D. Farmer, ``Long memory of the efficient market,'' {\em
  Nonlinear Dynamics and Econometrics}~{\bf 8(3)}, p.~1, 2004.

\bibitem{Lillo05}
F.~Lillo and J.~D. Farmer, ``The key role of liquidity fluctuations in
  determining the large price changes,'' {\em Fluctuations and noise
  letters}~{\bf 5(2)}, p.~L209, 2005.

\bibitem{Weber05}
P.~Weber and B.~Rosenow, ``Order book approach to price impact,'' {\em
  Quantitative Finance}~{\bf 5(4)}, p.~357, 2005.

\bibitem{Weber06}
P.~Weber and B.~Rosenow, ``Large stock price changes: volume or liquidity?,''
  {\em Quantitative Finance}~{\bf 6(1)}, p.~7, 2006.

\bibitem{Bouchaud06}
M.~Wyart, J.-P. Bouchaud, J.~Kockelkoren, M.~Potters, and M.~Vettorazzo,
  ``Relation between bid-ask spread, impact and volatility in double auction
  market,'' {\em preprint: physics/0603084} , 2006.

\bibitem{Cao03}
C.~Cao, O.~Hansch, and X.~Wang, ``The informational content of an open limit
  order book,'' {\em Working paper Pennsylvania State University} , 2003.

\bibitem{Alvarez91}
F.~Alvarez, A.~Alegria, and J.~Colmenero, ``Relationship between the
  time-domain {K}ohlrausch-{W}illiams-{W}atts and frequncy-domain
  havriliak-negami relaxation function,'' {\em Physical Review B}~{\bf 44},
  p.~7306, 1991.

\bibitem{Bouchaud02}
J.-P. Bouchaud, M.~M$\acute{\rm e}$zard, and M.~Potters, ``Statistical
  properties of stock order books: empirical results and models,'' {\em
  Quantitative Finance}~{\bf 2(4)}, pp.~251 -- 256, 2002.

\bibitem{Weron96}
K.~Weron and M.~Kotulski, ``On the {Cole-Cole} relaxation function and related
  {M}ittag-{L}effler distribution,'' {\em Physica A}~{\bf 232}, pp.~180--188,
  1996.

\bibitem{Magdziarz06}
M.~Magdziarz and K.~Weron, ``Anomalous diffusion schemes underlying the
  {C}ole-{C}ole relaxation: the role of the inverse-time $\alpha$-stable
  subordinator,'' {\em Physica A}~{\bf 367}, pp.~1--6, 2006.

\bibitem{Klafter86}
J.~Klafter and M.~F. Shlesinger, ``On the relationship among three theories of
  relaxation in disordered systems,'' {\em Proceedings of the National Academy
  of Sciences USA}~{\bf 83}, pp.~848--851, 1986.

\bibitem{Castaing90}
B.~Castaing, Y.~Gagne, and E.~J. Hopfinger, ``Velocity probability density
  functions of high {R}eynolds number turbulence,'' {\em Phys. D}~{\bf 46}(2),
  pp.~177--200, 1990.

\bibitem{Beck03}
C.~Beck, ``Superstatistics,'' {\em Physica A}~{\bf 322}, p.~267, 2003.

\bibitem{Honerkamp98}
J.~Honerkamp, {\em Statistical physics: an advanced approach with
  applications}, Springer-Verlag, Berlin, 1998.

\bibitem{Paul99}
W.~Paul and J.~Baschnagel, {\em Stochastic processes: from physics to finance},
  Springer, Berlin, 1999.

\bibitem{Barabasi05}
A.-L. Barab$\acute{\rm a}$si, ``The origin of bursts and heavy tails in human
  activity,'' {\em Nature}~{\bf 435}, p.~207, 2005.

\bibitem{Bartolozzi05}
M.~Bartolozzi, D.~B. Leinweber, and A.~W. Thomas, ``Stochastic opinion
  formation in scale-free networks,'' {\em Physical Review E}~{\bf 72},
  p.~046113, 2005.

\bibitem{Goh06}
K.-I. Goh and A.-L. Barab$\acute{\rm a}$si, ``Burstiness and memory in complex
  systems,'' {\em preprint: physics/0610233} , 2006.

\bibitem{Cont00}
R.~Cont and J.-P. Bouchaud, ``Herd behaviour and aggregate fluctuations in
  financial markets,'' {\em Macroeconomics Dynamics}~{\bf 4}, p.~170, 2000.

\bibitem{Bartolozzi04}
M.~Bartolozzi and A.~W. Thomas, ``Stochastic cellular automata model for stock
  market dynamics,'' {\em Physical Review E}~{\bf 69}, p.~046112, 2004.

\bibitem{Sornette04}
D.~Sornette, {\em Critical phenomena in natural sciences}, Springer-Verlag,
  Berlin, 2004.

\end{thebibliography}
\bibliographystyle{spiebib}

\end{document}